\documentclass[letter] {aa} 
\usepackage{amsmath}
\usepackage{graphicx}
\usepackage{color}
\usepackage{txfonts}
\usepackage[authoryear]{natbib}

\begin{document}
\title{Discovery of a flux-related change of the cyclotron line energy in \\
Her X-1}

\author{R.~Staubert\inst{1}, N.I.~Shakura\inst{2}, K.~Postnov\inst{2}, 
J.~Wilms\inst{3},  R.E.~Rothschild\inst{4}, W.~Coburn\inst{5},
L.~Rodina\inst{1}, D.~Klochkov\inst{1}}

\offprints{staubert@astro.uni-tuebingen.de}

\institute{
	Institut f\"ur Astronomie und Astrophysik, Universit\"at T\"ubingen,
	Sand 1, 72076 T\"ubingen, Germany
\and
	Sternberg Astronomical Institute, 119992, Moscow, Russia
\and
        Dr.\ Remeis Sternwarte, Astronomisches Institut der
	Universit\"at Erlangen-N\"urnberg, Sternwartstr. 7, 96049 Bamberg, Germany
\and
        Center for Astrophysics and Space Sciences, University of
        California at San Diego, La Jolla, CA 92093-0424, USA
\and
        Space Sciences Laboratory, University of California, Berkeley,
        CA 94720-7450, USA 
}

\date{}
\authorrunning{Staubert et al.}
\titlerunning{Cyclotron line energy changes in Her X-1}

\abstract
   {}
   {
We present the results of ten years of repeated measurements of the 
Cyclotron Resonance Scattering Feature (CRSF) in the spectrum of the 
binary X-ray pulsar \object{Her~X-1} and report the discovery of a positive 
correlation of the centroid energy of this absorption feature 
in pulse phase averaged spectra with source luminosity.
   }
   {
Our results are based on a uniform analysis of observations by
the \textsl{RXTE} satellite from 1996 to 2005, using sufficiently
long observations of 12 individual 35-day Main-On states of the 
source.
   }
   {
The mean centroid energy $E_c$ of the CRSF in pulse phase averaged 
spectra of Her~X-1 during this time is around 40\,keV, with 
significant variations from one Main-On state to the next. We find that 
the centroid energy of the CRSF in Her~X-1 changes by 
$\sim$ 5\% in energy for a factor of 2 in luminosity. The correlation 
is positive, contrary to what is observed in some high luminosity 
transient pulsars.  
   }
   {
Our finding is the first significant measurement of a positive
correlation between $E_c$ and luminosity in any X-ray pulsar. We
suggest that this behaviour is expected in the case of sub-Eddington 
accretion and present a calculation of a quantitative estimate, 
which is very consistent with the effect observed in Her~X-1. 
We urge that Her~X-1 is regularly monitored further and that 
other X-ray pulsars are investigated for a similar behaviour.  
   }
  {}

   
   \maketitle
%
%_______________________________________________________________

\section{Introduction}

The X-ray spectrum of the accreting binary pulsar Her~X-1 is
characterized by a power law continuum with exponential cut-off and an
apparent line-like feature, which was discovered in 1975
\citep{Truemper78}.
This feature is now generally accepted as an absorption feature around
40\,{\rm keV} due to resonant scattering of photons off electrons on
quantized energy levels (Landau levels) in the Teragauss magnetic
field at the polar cap of the neutron star. The feature is therefore
often referred to as a cyclotron resonant scattering feature (CRSF). The
energy spacing between the Landau levels is given by $E_\text{c} =
\hbar eB/m_{\rm e}c = 11.6\,\text{keV}\,B_{12}$, where
$B_{12}=B/10^{12}\,\text{G}$, providing a direct method of measuring
the magnetic field strength at the site of the emission of the X-ray
spectrum. The observed line energy is subject to gravitational
redshift, $z$, such that the magnetic field may be estimated by
$B_{12} = (1+z) E_\text{obs}/11.6\,{\rm keV}$.  The discovery of the
cyclotron feature in the spectrum of Her X-1 provided the first ever
`direct measurement' of the magnetic field strength of a neutron star,
in the sense that no other model assumptions are needed.  Originally
considered an exception, cyclotron features are now known to be rather
common in accreting X-ray pulsars, with more than a dozen binary
pulsars now being confirmed cyclotron line sources \citep{Coburn02}.  
In several objects, multiple lines have been detected (up to four 
harmonics, see \citealt{Staubert03}, \citealt{Heindl04} for reviews). \\ 
\indent In this \textsl{Letter} we present new results (from the last ten
years) on the energy of the cyclotron resonance scattering feature in the 
pulse averaged X-ray spectrum of Her~X-1, we summarize our knowledge 
about its variability, and report the discovery of a positive correlation
between $E_c$ and X-ray luminosity and offer a physical
explanation. A quantitative description of the physics of the accretion 
column under the condition of sub-Eddington accretion is consistent
with the observation. Preliminary results were given in \citet{Staubert06_mos}.

\section{Observations}

Her~X-1 is probably the best observed accreting binary X-ray pulsar.
Its X-ray spectrum, including the CRSF, has been measured by many
instruments since its discovery in 1975 \citep{Truemper78}. 
Here we report on a coherent analysis of ten years (1996--2005) of 
repeated observations by \textsl{RXTE}, including a reanalysis of those data 
already used by \citet{Gruber01}. Main-On state (35\,day-phases 
$<0.18$) observations (where the X-ray flux is the highest) 
of $>10$\,ks duration were used. These 
\textsl{RXTE} measurements are of high quality and yield comparatively 
small uncertainties in the measured quantities. The joint spectral 
analysis of \textsl{PCA} and \textsl{HEXTE} data was performed 
using the standard \textsl{XSPEC/FTOOLS} (6.0.4) software. 
For the spectral model we have chosen the \texttt{highecut}
model which is based on a power law continuum with exponential
cut-off, and the CRSF is modeled by an absorption line with a Gaussian 
optical depth profile. Details of the fitting procedure can be found in 
\citet{Coburn02}. Here we discuss results from spectral analysis of pulse 
phase averaged spectra only (a pulse phase resolved analysis is in
progress). Table~1 summarizes the observation
dates, the measured CRSF centroid energies and the maximum X-ray fluxes
for the corresponding 35\,day Main-On states, as measured by the 
\textsl{RXTE/ASM} (formally determined by fitting the flux history 
of the respective Main-On by a template function derived from the 
average of many Main-On cycles). Table~1 also lists an additional data 
point from an observation with \textsl{INTEGRAL} \citep{Klochkov06}. 

%Table 1 -------------------------------------------------------------------
\begin{table}
\caption[]{Recent cyclotron line energy measurements in Her~X-1 by
  \textsl{RXTE} and \textsl{INTEGRAL}. Uncertainties are at the 68\% level.
The data point from \textsl{INTEGRAL} is from \citet{Klochkov06}. 35\,d 
cycle numbering is according to \citet{Staubert83}.}
\begin{center}
\begin{tabular}{lllll}
\hline\noalign{\smallskip}
Observation & 35\,d & Center    & Line Energy     & max. Flux \\
month/year  & cycle & MJD       & keV             & ASM cts/s \\
\hline\noalign{\smallskip}
\textsl{RXTE} \\       
July 96     & 257   & 50029.75  & $41.12\pm0.55$  & $7.37\pm0.34$ \\
Sep 97      & 269   & 50707.06  & $40.62\pm0.49$  & $7.49\pm0.73$ \\
Dec 00      & 304   & 51897.69  & $40.07\pm0.31$  & $6.04\pm0.47$ \\
Jan 01      & 305   & 51933.67  & $39.05\pm0.55$  & $5.72\pm0.34$ \\
May 01      & 308   & 52035.48  & $39.93\pm0.63$  & $7.15\pm0.50$ \\
June 01     & 309   & 52071.16  & $39.73\pm0.52$  & $6.93\pm0.20$ \\
Dec 01      & 314   & 52245.09  & $40.04\pm0.22$  & \\
Aug 02      & 321   & 52492.96  & $40.01\pm0.29$  & $7.19\pm0.26$ \\
Nov 02      & 324   & 52599.32  & $40.51\pm0.13$  & $7.64\pm0.30$ \\
Dec 02      & 325   & 52634.01  & $40.60\pm0.41$  & $7.55\pm0.34$ \\
Oct 04      & 344   & 53300.95  & $38.51\pm0.51$  & $4.50\pm0.24$ \\
July 05     & 352   & 53577.35  & $38.95\pm0.52$  & $5.12\pm0.37$ \\
\textsl{INTEGRAL} \\   
July 05      & 352 & 53576.00  & $38.50\pm0.70$  & $5.12\pm0.37$ \\
\noalign{\smallskip}\hline
\end{tabular}\end{center}
\end{table}
%Table 1 -------------------------------------------------------------------

\section{Results}

Fig.~1 displays the measured CRSF centroid energies as a function of 
time, combining the results of our analysis with historical measurements 
before the \textsl{RXTE} era, as taken from the compilation by 
\citet[their tables~2 and~3]{Gruber01}, as well as the 2005 measurement 
by \textsl{INTEGRAL} \citep{Klochkov06}. Also shown in Fig.~1 (right hand
scale) is the maximum flux of the corresponding 35\,d Main-On states as measured 
by \textsl{RXTE/ASM}. We report on two main results. 

Firstly, we confirm the apparent difference in the mean cyclotron line energy 
before and after 1991, first pointed out by \citet{Gruber01}.
Taking the measured values of $E_c$ and their stated uncertainties
at face value, the mean cyclotron line energies $\langle E_c \rangle$ 
(dotted lines in Fig.~1) from 
all measurements before 1991 is $34.9\pm0.3\,\text{keV}$, the corresponding 
value for all measurements after 1991 is $40.3\pm0.1\,\text{keV}$ 
($40.2\pm0.1\,\text{keV}$ for \textsl{RXTE} results only, showing that
the very high value measured by \textsl{BATSE} is not decisive). However, a 
comparison of measurements from different instruments is difficult because of 
systematic uncertainties due to calibration and analysis techniques. 
Nevertheless, we believe that the large difference of $\sim 5\,\text{keV}$ 
between the mean values and the relative internal consistency within the
two groups (5 different instruments before 1991 and four after 1991) most 
likely indicate real physics (we will comment on this in the discussion). 

%Fig. 1 -------------------------------------------------------------------
\begin{figure}
\resizebox{\hsize}{!}{\includegraphics{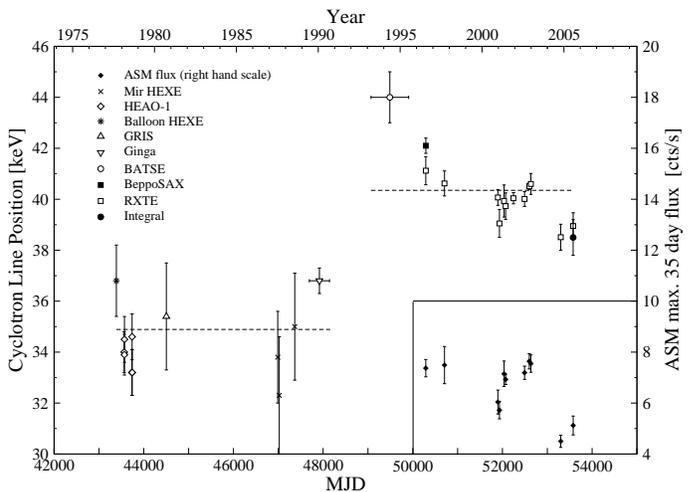}}
\caption{The centroid energy of the phase averaged cyclotron resonance
  line feature in Her~X-1 since its discovery. Data from before 1997
  were originally compiled by \citet{Gruber01}, where the original 
  references can be found. The results of our reanalysis of RXTE data 
  are consistent with the earlier results.
  The \textsl{INTEGRAL} point is from \citet{Klochkov06}.
  The inset at the lower right corner shows the ASM maximum 35\,d flux
  (right hand scale). 
  The dotted lines are mean cyclotron line energies 
  $\langle E_c \rangle$ before and after 1991.
}
\end{figure}
%Fig. 1 -------------------------------------------------------------------

Secondly, we present a new result: we have detected a positive correlation
between the maximum 35\,d  Main-On state X-ray flux ($2-10\,\text{keV}$), 
which we take as a measure of the X-ray luminosity, and the cyclotron line 
centroid energy $E_c$. If we take the locally measured fluxes (e.g. from 
our spectral fits) we find that they track the maximum ASM flux quite
well and they show the same correlation with $E_c$. However, we prefer
the Main-On state maximum flux as the better information on the accretion 
state of the source, since the locally measured flux is subject to modulation
by variable absorption/shading by the accretion disk.

For a quantitative analysis of this correlation, which is already evident 
from Fig.~1, we restrict ourselves to our results from \textsl{RXTE} data only 
(see Table~1), because they offer the advantage of maximum comparability due 
to a uniform analysis of data from the same instruments. In Fig.~2 we plot 
the cyclotron line centroid energy $E_c$ as a function of the maximum 35\,d 
Main-On state \textsl{RXTE/ASM} flux.
The dashed line is from a linear fit, taking uncertainties of both variables
into account (see e.g. \textsl{Numerical Recipies, 15.3}), and it
defines a slope of $0.66\pm0.10\,(68\%)\,\text{keV/(ASM cts/s})$. 
Pearsons linear correlation coefficient is 0.90, the Spearman Rank 
correlation coefficient is 0.85 
(corresponding to the probabilities to find these correlations by
chance of $\leq 6~10^{-5}$ and $\leq 4~10^{-4}$, respectively. We consider this 
correlation as highly significant.

We also point out here, that the apparent decrease of $E_c$ (and
the correlated \textsl{ASM} count rate) with time after 1991, seen 
in Fig.~1, is an artifact due to the general variability of the 
source and the random way by which the data happened to be taken,
and not a secular decrease. Despite strong short-term variability 
the \textsl{RXTE/ASM} light curve of Her~X-1 does not provide any indication 
for a long-term ($\geq 5$\,yrs) change of the mean luminosity.

\section{Variations in cyclotron lines}

Apart from variations with pulse phase, shifts of line positions have
in phase averaged spectra been found in correlation with changing 
X-ray luminosity in a number of high luminosity transient sources. 
However, these correlations are negative: $E_c$ is reduced when $L_x$ 
increases. \citet{Mihara98} have interpreted this for \object{4U~0115+63}, 
\object{Cep~X-4}, and \object{V~0332+53}, 
as due to a change in height of the shock (and emission) region above the 
surface of the neutron star with changing mass accretion rate, $\dot{M}$. 
The accretion is in the super-Eddington regime.
In the model of \citet{Burnard91}, the height of the polar accretion structure 
is tied strongly to $\dot{M}$. From this model one expects that an increase
in accretion rate leads to an increase in the height of the scattering 
region above the neutron star surface, and therfore to a decrease in magnetic field
strength and hence $E_c$. During the 2004/2005 outburst of V~0332+53 
a clear anti-correlation of the line position with X-ray flux was observed 
\citep{Kreyken05, Tsygankov06a, Mowlavi06}. Analysing \textsl{RXTE} data of 
the Feb--Apr 1999 outburst of 4U~0115+63,  
both \citet{Nakajima06} and \citet{Tsygankov06b} find a general 
anti-correlation between $E_c$ and luminosity. There seems to be a steep
dependence in a limited luminosity range ($2...5\cdot10^{37}$\,erg/s), and
neighboring regions at lower and higher luminosities with near independence 
of the two quantities. At the lowest luminosity the data are even
consistent with a reversal of the dependence (albeit with low statistical
significance).

%Fig.2 -------------------------------------------------------------------
\begin{figure}
\resizebox{\hsize}{!}{\includegraphics{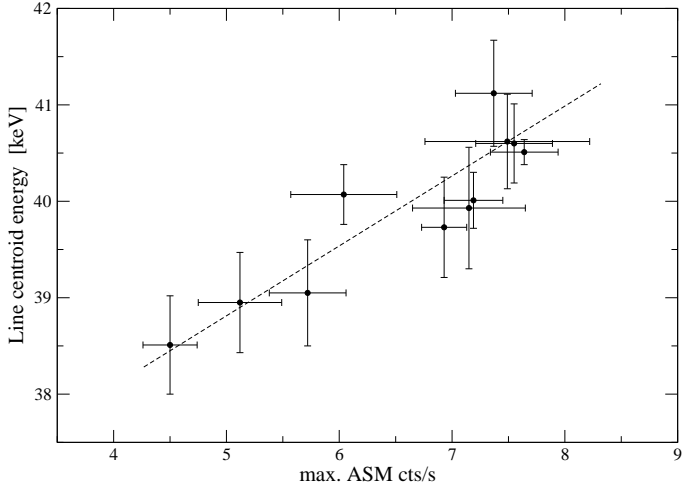}}
\caption{The centroid cyclotron line energy of Her~X-1 versus the 
maximum flux during the corresponding 35\,day Main-On as observed by the 
\textsl{RXTE/ASM}. A linear fit (taking uncertainties in both variables into 
account) defines a slope of $0.66\pm0.10\,\text{keV/(ASM~cts/s)}$.}
\end{figure}
%Fig.2 -------------------------------------------------------------------

For Her~X-1, however, so far no clear relation between the cyclotron 
line energy $E_c$ and X-ray flux of any kind had been found 
\citep[see e.g.][]{Mihara98, Gruber01} (but see our note in the Discussion). 
In this Letter we report the first detection of a clear positive correlation 
between $E_c$ and X-ray luminosity in Her~X-1 (and in any
X-ray pulsar, if one requires a convincing evidence of sufficient 
significance; see the Discussion).

\section{Physics in the sub-Eddington accretion regime}

In the following we suggest a physical explanation of our
finding and attempt a quantitative estimate of the expected effect.
Since the pioneering paper by \citet{BaskoSunyaev76}, it has been known 
that the structure of the accretion column near the neutron star surface 
in X-ray pulsars differs for high-luminosity ($L>L_c$) and
low-luminosity ($L<L_c$) regimes, where the critical luminosity $L_c$
%$L_*\sim 10^{37}$ erg/s is the 
is given by the local Eddington luminosity $L_E$ \citep{Nelson93} 

\vspace{-2mm}
\begin{equation}
L_E=\frac{2\pi \text{GMcm}_\text{p}}{\sigma_T}\left(\frac{\sigma_T}{\sigma_m} \right)
\theta_c^2 
\simeq 10^{36}~\text{erg/s}~\left(\frac{\sigma_T}{\sigma_m} \right)
\left(\frac{\theta_c}{0.1}\right)^2
\end{equation}

Here $M$ is the NS mass (assumed to be 1.4 $M_\odot$ in the numerical
estimation), $\sigma_m$ is the photon-electron scattering
cross-section in the magnetic field, $\sigma_T$ is the Thomson
cross-section, and $\theta_c$  is the half-opening angle of the polar cap
magnetic field lines. 
We note that sub-/super-Eddington accretion cannot be judged by the 
observed luminosity alone, decisive is the local accretion rate which depends 
on the accretion area and the photon-electron scattering cross section, 
which in turn depends on the magnetic field (strength and structure).
In the radiation-dominated super-Eddington regime 
the height of the accretion column increases with luminosity \citep{Burnard91}. 
As discussed above, this explains the observed anti-correlation between
the CRSF energy in spectra of bright transient X-ray pulsars with luminosity
\citep{Mihara98, Tsygankov06a, Tsygankov06b,Nakajima06}, as the magnetic field 
value decreases with the height above the NS surface.  

In the 
sub-Eddington regime of accretion, however,
we can expect a different behaviour of the CRSF centroid energy
when the luminosity changes, turning even to the opposite to what is 
observed in high-luminosity X-ray pulsars. For the set of
\textsl{RXTE} observations of Her~X-1 we have a 
case of sub-Eddington accretion in this X-ray pulsar. With the cyclotron 
line energy $E_{\rm c}$ of about 40\,keV, the
cross-section of photon--electron scattering in the direction along 
the magnetic field lines (which is of interest for the radiation 
pressure against the infalling material) decreases $\propto (E/E_c)^2$ 
\citep[][and references therein]{HardingLai06}. At low energies where 
most of the photons are produced in the case of Her X-1, 
$(\sigma_m/\sigma_T)\ll1$, and for an X-ray luminosity of a few times 
$10^{37}$ erg/s and $\theta_c$ not much smaller than 0.1 
($\sim 6^\circ$) the source is in the sub-Eddington regime.

The physical picture of magnetic accretion in low-luminosity X-ray pulsars 
was discussed by \citet{Nelson93}. We assume that the accreting 
protons lose their kinetic energy in an electron-proton atmosphere due to 
the Coulomb drag and collective plasma effects and neglect the possibility
for the formation of a collisionless shock \citep[see e.g.][]{LangerRappaport82}. 
The characteristic braking length for protons can be identified with their
mean free path $l_*\sim 1/n_e\sigma$, where $\sigma$ is the effective 
interaction cross-section. 
A change in the observed cyclotron line energy can be associated with
a change in the height of the emission region above the neutron star surface. 
We write the observed cyclotron energy as 

\vspace{-1mm}
\begin{equation}
E_c=\left(\frac{\hbar eB}{m_ec}\right)g_{00}^{1/2} 
\end{equation}

where $B$ is the polar magnetic field strength, 
$g_{00}^{1/2}=\sqrt{1-R_g/r}=1/(1+z)$ 
is the Schwarzschild metric coefficient determining the gravitational 
redshift $z$. $R_g=2GM/c^2$ is the gravitational radius.
Identifying the characteristic length $l_*$ with the height of the 
scattering region above the surface of the NS, we define $r=R+l_*$
(with $R$ being the radius of the NS). Then the fractional change of the 
cyclotron line energy is
$\Delta E_c/E_c = \Delta B/B +(1/2) (\Delta g_{00}/g_{00})$.
Assuming a pure dipole field ($B \propto r^{-3}$) we obtain 
$\Delta E_c/E_c = -3 \Delta r/R + (1/2) (R_g/R) (\Delta r/R)/g_{00}$.

The redshift term can be neglected since for a canonical neutron star
its contribution is only of the order of 10\% of the dipole term.
Making the identification $\Delta r/l_*=-\Delta n_e/n_e$, we arrive at 

\begin{equation}
\frac{\Delta E_c}{E_c}=3\frac{l_*}{R}\frac{\Delta n_e}{n_e}\
\end{equation}

Now we need to find how the density in the region of energy 
release and line formation is related to the luminosity. There is 
no simple estimate of this, but we can use the approach by 
\citet{Nelson93} to find the structure of the accretion mound.  
In addition to the hydrostatic term considered by \citet{Nelson93}, we add 
the dynamical pressure of infalling protons so that the total pressure is 
$P=2n_ekT=gy + (\rho_0v_0^2-\rho v^2)$.
Here $y$ is the mass column density, $T$ is the temperature, 
$\rho_0$ and $v_0$ are the density and velocity of the free-falling
matter at the beginning of the braking region. It is of course not a 
self-consistent treatment of the problem, but a step forward to
account for dynamical pressure of accreting protons inside the braking 
region. For the free fall acceleration near the NS surface 
$g\simeq  2\cdot 10^{14}\,\text{cm~s}^{-2}$ and 
$y\simeq 20\,\text{g~cm}^{-2}$ \citep{Miller87}, the hydrostatic term 
is not higher than $4\cdot 10^{15}\,\text{dyn~cm}^{-2}$. 
The dynamical term can be expressed using the solution found by 
\citet{Nelson93} and the continuity equation $\dot M=\rho v A$ 
($A$ is the accretion area for one pole) in the form  
$\rho_0v_0^2-\rho v^2= (\dot M/A) v_0(1-(1-\tau/\tau_*)^{1/4})$, where
$\tau$ is the Thompson optical depth and $\tau_*$ is the proton 
stopping depth. Clearly, the dynamical term vanishes at the base of 
the accretion column (i.e. at $\tau=\tau_*$) where the bulk kinetic
energy of the protons is transformed into their thermal motion. 
It dominates, however, at the beginning of the deceleration region
(for small $y$ and small $\tau$, and for typical values like 
$\dot M=10^{17}\,\text{g~s}^{-1}$, $A=1/400$ of the NS surface area  
and $v_0=10^{10}\,\text{cm~s}^{-1})$.    
As the stopping depth is $\tau_*\sim 50$ for magnetic accretion 
\citep{Nelson93}, we find for $\tau\ll \tau_*$ (from where the 
emission escapes)
$n_e\simeq (\dot M/A) (v_0/2kT) (\tau/4\tau_*)$.

As long as $l_*\ll R$ we can neglect a variation in $A$ (this is 
justified since the stopping length for $\tau_* \sim 50$ and 
$n_e\sim 10^{24}\,\text{cm}^{-3}$ is very small), so main parameter 
which determines the density inside the stopping region is $\dot M$. 
We also neglect the principle dependence of $A$ on $\dot M$
(due to the change of the disks inner radius $R_{in}$) since this
dependence is weak (for dipolar field $R_{in} \propto \dot M^{-2/7}$).
So we finally obtain:

\begin{equation}
\frac{\Delta E_c}{E_c}=3\frac{l_*}{R}\frac{\Delta  \dot M}{\dot M}=3\frac{l_*}{R}\frac{\Delta  L}{L}\
\end{equation}

We see that in the sub-Eddington accretion regime the fractional
change in cyclotron line energy can be directly proportional to 
the fractional change in luminosity. For example, for the observed 
luminosity variations in Her X-1 $\Delta L/L\sim 1$ 
(a change of a factor of two in the observed flux) the formula 
above gives $\Delta E_c/E_c\sim 3 l_*/R$. The stopping length is 
$l_*=\tau_*/(n_e\sigma_T)$, the density estimation for 
$\dot M=2\cdot 10^{17}\,\text{g~s}^{-1}$, $v_0=10^{10}\,\text{cm~s}^{-1}$, 
$kT=10\,\text{keV}$, and $\tau\sim 1$ yields $n_e\sim 10^{22}\,\text{cm}^{-3}$, 
so $l_*\sim 10^4\,\text{cm}$. 
So the expected fractional change in the cyclotron line energy is 
$\Delta E_c/E_c \simeq 0.03$, which is very close to the observed 
value of 5\%. We note, however, that we used a fiducial value  of 
$\tau_*=50$ appropriate for a weak magnetic field \citep{Nelson93}, but 
in strong magnetic fields this value can be higher, the electron 
density would decrease and $l_*$ increase, accordingly. So we suggest 
that a measurement of $\Delta E_c/E_c$ can in fact be used to assess 
the stopping length of protons in the regime of sub-Eddington magnetic 
accretion onto neutron stars.  

\section{Discussion}

Using standard parameters the calculations above for the sub-Eddington
accretion regime lead to an estimate which is close to the observed effect 
in Her~X-1 ($\sim$ 5\% in $E_c$ for a factor of two in luminosity).
The decisive parameter is the local Eddington rate at the neutron star. 
Its value depends on the area upon which accretion proceeds, so it is
expected to vary from pulsar to pulsar. In transient pulsars, such as 
4U~0115+63, we may already have evidence for a transition from 
super- to sub-Eddington accretion \citep{Tsygankov06b,Nakajima06,Terada06} 
at the decline of the outburst when the luminosity drops below 
$\sim 5 \cdot 10^{37}$\,erg/s. The sharp jump in the observed cyclotron 
line energy in 4U~0115+63 reported by \citet{Tsygankov06b} could be
a sign of an abrupt change in the structure of the accretion column
during such a transition.

It is conceivable that long-term variations in the accretion rate exist 
also in the persistent X-ray pulsar Her X-1. If so, 
we speculate that the abrupt jump in the cyclotron line energy in 
Her~X-1 noticed in the early 1990s (Fig.~1) might be associated with
a transition of this pulsar from super-Eddington to sub-Eddington
accretion. We plan to have a close look at the historical data with
respect to variations in absolute flux (which is not an easy task
because of the difficulties connected with the inter-calibration of
different instruments). Taking the binary X-ray pulsar \object{A~0535+262} 
as an example, the absence of a strong luminosity dependence reported by 
\citet{Terada06} may indicate that this source stays always in the 
sub-Eddington regime (for example, because the value $l_*$ increases with 
the magnetic field strength). If anything, we read a slight 
increase in $E_c$ with luminosity from Fig.~4 of \citet{Terada06}. 
Another indication of a positive correlation between $E_c$ and luminosity 
(and a qualitative interpretation) has been reported by 
\citet{LaBarbera05} for \object{GX~301--2} (albeit only on the basis of 
two observations and at a significance level of about 2
standard deviations).
While \citet{Mihara98} and \citet{Gruber01} have shown that Her~X-1 does
not follow the negative correlation known from other sources, their data
(Table~2 and Fig.~5, respectively) do in fact both show a small positive 
trend (not noticed by the two groups of authors at the time). 
While the statistical significance is low, both sets of data are 
consistent with our finding of 5\% increase in $E_c$ for a factor of two 
increase in flux.

Finally, we note that Her~X-1 is the only highly magnetized accreting 
pulsar for which repeated observations over longer periods of time exist. We
therefore urge that the source continues to be monitored regularly, the
source is bright and persistent and gives excellent statistics in relatively
short observations.

\begin{acknowledgements}
This work was supported by DFG through grants Sta 173/31-1,2 and
436~RUS~113/717 and RFBR grants RFFI-NNIO-03-02-04003 and 06-02-16025. 
We thank the anonymous referee for very good suggestions.
\end{acknowledgements}

\bibliographystyle{aa}
\bibliography{7098ref}

\end{document}